\documentclass[a4paper,11pt]{article}
\pdfoutput=1 

\usepackage{jcappub} 

\usepackage[T1]{fontenc} 
\usepackage{graphicx}
\usepackage{latexsym}
\usepackage{amssymb,amsmath}

\def\l{\left}
\def\r{\right}
\renewcommand{\a}{\alpha}
\renewcommand{\b}{\beta}
\renewcommand{\d}{\delta}

\newcommand{\g}{\gamma}

\newcommand{\la}{\lambda}

\newcommand{\pa}{\partial}
\newcommand{\td}{\textrm{d}}
\newcommand{\nn}{\nonumber\\}

\newcommand{\LM}{\mathcal{L}}
\newcommand{\HM}{\mathcal{H}}
\newcommand{\RM}{\mathcal{R}}
\def\be{\begin{equation}}
\def\ee{\end{equation}}
\def\bea{\begin{eqnarray}}
\def\eea{\end{eqnarray}}
\def\bal{\begin{align}}
\def\eal{\end{align}}
\newcommand{\bit}{\begin{itemize}}
\newcommand{\eit}{\end{itemize}}

\title{Two-field mimetic gravity revisited and Hamiltonian analysis}

\author[a,b]{Liuyuan Shen}
\author[c,d]{Yunlong Zheng}
\author[e]{Mingzhe Li}

\affiliation[a]{Department of Physics, Nanjing University, Nanjing 210093, China}
\affiliation[b]{Joint Center for Particle, Nuclear Physics and Cosmology, Nanjing 210093, China}
\affiliation[c]{CAS Key Laboratory for Researches in Galaxies and Cosmology, Department of Astronomy, University of Science and Technology of China, Hefei, Anhui 230026, China}
\affiliation[d]{School of Astronomy and Space Science, University of Science and Technology of China, Hefei, Anhui 230026, China}
\affiliation[e]{Interdisciplinary Center for Theoretical Study, University of Science and Technology of China, Hefei, Anhui 230026, China}

\emailAdd{zhyunl@ustc.edu.cn}

\abstract{
We revisit the two-field mimetic gravity model with shift symmetries recently proposed in the literature, especially the problems of degrees of freedom and stabilities. We first study the model at the linear cosmological perturbation level by quadratic Lagrangian and Hamiltonian formulations. We show that there are actually two (instead of one) scalar degrees of freedom in this model in addition to two tensor modes.  We then push on the study to the full non-linear level in terms of the Hamiltonian analysis, and confirm our result from the linear perturbation theory.  We also consider the case where the kinetic terms of the two mimetic scalar fields have opposite signs in the constraint equation. We point out that in this case the model always suffers from the ghost instability problem.

}

\begin{document}
\hfill USTC-ICTS-19-23
\maketitle
\flushbottom

\section{Introduction}\label{sec:intro}
%
Mimetic scenario was  proposed by Chamseddine and Mukhanov \cite{Chamseddine:2013kea}  as a theory of  modifying  Einstein's general relativity. The idea is to express the physical metric $g_{\mu\nu}$ in the Einstein-Hilbert action with
\begin{equation}\label{conformal}
g_{\mu\nu}=\left(\tilde{g}^{\alpha\beta}\phi_{,\alpha}\phi_{,\beta}\right)\tilde{g}_{\mu\nu}
\end{equation}
by an auxiliary metric $\tilde{g}_{\mu\nu}$ and a covariant derivative of a scalar field, $\phi_{,\alpha}=\nabla_{\alpha}\phi$, so that it is invariant under Weyl rescalings of $\tilde{g}_{\mu\nu}$. With the physical metric, the scalar field satisfies the constraint:
\begin{equation}
g^{\mu\nu}\phi_{,\mu}\phi_{,\nu}=1~,
\end{equation}
as a component it can mimic the dark matter, hence the theory is dubbed the mimetic dark matter. 
Alternatively, the above mimetic constraint can be imposed in the action by  a Lagrange multiplier \cite{Golovnev:2013jxa}. So the action for the mimetic model takes the following form
\begin{equation}
S= \int\td^4 x \sqrt{-g} \left[\frac{1}{2} R+\lambda\left(g^{\mu\nu}\phi_{,\mu}\phi_{,\nu}-1 \right)+\mathcal{L} _{m}\right]~,
\end{equation}
where we have used the unit reduced Planck mass $M_p^2 = 1/(8\pi G) = 1$ and the most negative signature for the metric.
The mimetic model then was  generalized in Ref.~\cite{Chamseddine:2014vna,Lim:2010yk} by adding  a potential $V (\phi)$ to the mimetic field for phenomenological applications. 
It is shown that the generalized mimetic model  can provide inflation, bounce, dark energy, and so on with appropriate choice of the potential.  Therefore, it has attracted  extensive cosmological and astrophysical interests \cite{Saadi:2014jfa,Mirzagholi:2014ifa,Matsumoto:2015wja,Ramazanov:2015pha,Myrzakulov:2015kda, Astashenok:2015qzw,Chamseddine:2016uyr,Nojiri:2016vhu,Babichev:2016jzg,Chamseddine:2016ktu,Shen:2017rya,Abbassi:2018ywq,Casalino:2018tcd,Vagnozzi:2017ilo,Dutta:2017fjw,Chamseddine:2019pux}. 
 Meanwhile, mimetic scenario has been applied in various modified gravity theories  
\cite{Nojiri:2014zqa,Leon:2014yua,Astashenok:2015haa,Myrzakulov:2015qaa,Rabochaya:2015haa,Arroja:2015yvd,Cognola:2016gjy,Nojiri:2016ppu,Momeni:2014qta,Chamseddine:2018qym,Chamseddine:2018gqh,Zhong:2017uhn,Zhong:2018fdq,Chamseddine:2019gjh,Liu:2017puc}.
 The Hamiltonian analysis of various mimetic models have also been investigated in Refs.~\cite{Malaeb:2014vua,Chaichian:2014qba,Takahashi:2017pje,Zheng:2018cuc,Ganz:2018mqi,Malaeb:2019rdl,Ganz:2019vre}. Moreover, there are also some other theoretic developments  \cite{Hammer:2015pcx,Golovnev:2018icm,Langlois:2018jdg,Firouzjahi:2018xob,Gorji:2018okn}.
 See Ref.~\cite{Sebastiani:2016ras} for a review.   

Although the the original mimetic gravity is free of pathologies, the perturbation analysis \cite{Chamseddine:2014vna} shows that the fluctuation is non-propagating even in the existence of a potential, thus can't be quantized in a usual way. To have a propagating scalar mode, higher derivative terms of the mimetic field has been introduced to the action \cite{Chamseddine:2014vna}. Unfortunately, such modifications suffer from ghost or gradient instabilities \cite{Ijjas:2016pad,Firouzjahi:2017txv}. Then it was suggested \cite{Zheng:2017qfs,Hirano:2017zox,Gorji:2017cai} to overcome this difficulty by introducing the direct couplings of the higher derivatives of the mimetic field to the curvature of the spacetime.

One interesting point about mimetic gravity is its close relation with non-invertible transformation. It has been shown in Refs.~\cite{Deruelle:2014zza,Arroja:2015wpa,Domenech:2015tca} that mimetic gravity can be obtained through a non-invertible disformal transformation \cite{Bekenstein:1992pj} where the degrees of freedom (DOFs) are no longer preserved \footnote{The equivalence between two frames connected by invertible transformation is established, see e.g. Ref.~\cite{Takahashi:2017zgr}.} between the two frames and the additional DOF play the role of dark matter. Following this idea, the two-field extension of the mimetic scenario \cite{Firouzjahi:2018xob}
 was recently  proposed by looking at the singular  limit of conformal transformation 
 \begin{equation}\label{twoconformal}
g_{\mu\nu}=A(\phi,\psi,\tilde{X},\tilde{Y},\tilde{Z})\tilde{g}_{\mu\nu}.
\end{equation}
where  $\tilde{X} \equiv \tilde{g}^{\mu \nu} \phi_{, \mu} \phi_{, \nu},~ \tilde{Y} \equiv \tilde{g}^{\mu \nu} \psi_{, \mu} \psi_{, \nu}$ and $\tilde{Z} \equiv \tilde{g}^{\mu \nu} \phi_{, \mu} \psi_{, \nu}.$ 
 \footnote{The notations in this paper are slightly different from those of Ref.~\cite{Firouzjahi:2018xob}.}
 The non-invertible condition is derived to be 
 \be\label{cd1}
	A(\phi,\psi,\tilde{X},\tilde{Y},\tilde{Z})=\tilde{X} A_{,\tilde{X}}+\tilde{Y} A_{,\tilde{Y}}+\tilde{Z} A_{,\tilde{Z}},
\ee
 and after imposing shift symmetries on $\phi$ and $\psi$, the nontrivial solution for $A$ is \cite{Firouzjahi:2018xob}
 \be\label{cd2}
     A=\a\tilde{X}+\b\tilde{Y},
 \ee
where $\a$ and $\b$ are  nonvanishing  constants. The corresponding mimetic constraint is given by $\a X+\b Y=1$. Interestingly, it was found \cite{Firouzjahi:2018xob} that this setup still mimics the dark matter at the cosmological background level. 

However, for perturbations, \cite{Firouzjahi:2018xob} claimed that the adiabatic mode vanishes, i.e., $\mathcal{R}=0$, while the entropy mode is healthy and propagates with the unity sound speed, then concluded that there is only one scalar DOF in the two-field mimetic model.

In this paper, we will revisit the problem of DOFs in the two-field mimetic gravity model with shift symmetries. First we will consider the theory of linear perturbations around the homogeneous cosmological background. By employing the quadratic Lagrangian formulation as well as the Hamiltonian formulation, we will show that there are actually two scalar DOFs (instead of one) in this model. This is different from the result of \cite{Firouzjahi:2018xob}. Then we will push on our study to the full non-linear level using the Hamiltonian analysis, and confirm our result obtained from linear perturbation theory.  In addition, we will also study the case where the kinetic terms of the two mimetic scalar fields have opposite signs in the constraint equation. This has not been considered in the literature. We will show that in this case the model always suffers from the problem of ghost instability.

This paper is organized as follows. In Sec.~\ref{sec:framework}, we give a brief review of the two-field mimetic gravity model. In Sec.~\ref{sec:cosmology}, we reanalyze the two field mimetic gravity at the linear perturbation level, to find out the perturbative degrees of freedom and analyze their stabilities. In Sec.~\ref{sec:Hamiltonian} .we perform the Hamiltonian analysis at the full non-linear level and check the result from linear perturbation theory. Finally, we conclude in Sec.~\ref{sec:discussion}.

\section{two-field mimetic gravity}
\label{sec:framework}
 As pointed out in Refs.~\cite{Deruelle:2014zza,Arroja:2015wpa},  the orignial single field mimetic gravity can be generated by the singular limit  of the conformal transformation. In this regard, one can extend the mimetic gravity to  two scalar fields  by considering the conformal transformation \eqref{twoconformal}
 and looking for its non-invertible condition \cite{Firouzjahi:2018xob}.  One can first look at the Jacobian of the transformation $\frac{\pa g_{\mu\nu}}{\pa \tilde{g}_{\a\b}}$ and require the  eigenvalue to  vanish.  Finally, the condition on $A$  for the singular limit of  conformal transformation \eqref{twoconformal} is found to be
\be\label{cd1}
	A(\phi,\psi,\tilde{X},\tilde{Y},\tilde{Z})=\tilde{X} A_{,\tilde{X}}+\tilde{Y} A_{,\tilde{Y}}+\tilde{Z} A_{,\tilde{Z}}~.
\ee
 We obtain
\be \label{cd2}
  A(\phi,\psi,c \tilde{X},c \tilde{Y},c \tilde{Z})=c   A(\phi,\psi,\tilde{X},\tilde{Y},\tilde{Z})~,
\ee
which means  $A$ is a homogeneous function of degree one with respect
to  $\tilde{X},\tilde{Y}$ and $\tilde{Z}$.
 One  nontrivial solution \cite{Firouzjahi:2018xob} for $A$ is \footnote{Other solutions inlcude, for example, $A=\frac{\tilde{X}^2}{\tilde{Y}}$, $A=\sqrt{\tilde{X} \tilde{Y}}$ and so on, which will be discussed in detail in the future work.}
 \be\label{cd2}
     A(\phi,\psi,\tilde{X},\tilde{Y},\tilde{Z})=\a(\phi,\psi) \tilde{X}+\b(\phi,\psi) \tilde{Y}+2\g(\phi,\psi) \tilde{Z},
 \ee
which is a linear function of $\tilde{X},\tilde{Y}$ and $\tilde{Z}$ at the same time. In this work, we only consider this special solution.

Without loss of generality, one can set $\g = 0$ in the analysis below as the cross term $\tilde{Z}$ can be removed by a linear transformation of the field space. The non-invertible conformal transformation takes the form
\begin{equation}\label{con2}
g_{\mu\nu}=(\a\tilde{X}+\b \tilde{Y})\tilde{g}_{\mu\nu}.
\end{equation}
One can easily obtain the corresponding  two-field mimetic constraint 
\be\label{mc}
 \a X+\b Y=1,
\ee
 which is similar to the single field case. Here $X,Y$ are defined by the physical metric as $X=g^{\mu\nu}\phi_{,\mu}\phi_{,\nu} \rm{~and~} Y=g^{\mu\nu}\psi_{,\mu}\psi_{,\nu}$. 
 
In this paper, we impose shift symmetries on both scalars $\phi$ and $\psi$ as has been done in Ref.~\cite{Firouzjahi:2018xob}, so that $\a$ and $\b$ are constants. The studies on general cases without shift symmetries are left for future publications.  The constants $\a$ and $\b$ can be further absorbed into the fields through the field redefinitions $\phi \rightarrow \phi / \sqrt{|\alpha|}$ and $\psi \rightarrow \psi / \sqrt{{|\beta|}}$,  so the non-invertible conformal transformation can be written as
\begin{equation}\label{con3}
g_{\mu\nu}=(\tilde{X}+ c~\tilde{Y})\tilde{g}_{\mu\nu}~,
\end{equation}
where $c=\pm1$. The appearance of $c$ depends on the signs of $\alpha/|\alpha|$ and $\beta/|\beta|$. But only the relative sign is important, so here we assume $\a$ is positive and leave the sign of $\beta$ free.  Ref.~\cite{Firouzjahi:2018xob} only considered the case $c=1$,\footnote{Actually this case was first considered in \cite{Vikman:2017gxs} where  a complex field is used instead of two real fields.} here we will consider both cases. In terms of Eq. (\ref{con3}), we obtain  
the corresponding  two-field mimetic constraint equation:
\be\label{mc2}
    X+c~Y=1.
\ee

Therefore, the action of the two-field mimetic gravity  is  the Einstein-Hilbert action constructed in terms of the  physical metric \eqref{con3} :
\be
S=\int\td^{4} x \sqrt{-g(\tilde{g}, \phi, \psi)}\frac{1}{2} R(g(\tilde{g}, \phi, \psi)).
\ee
For convenience, one can write the action in terms of a Lagrange multiplier, 
\be\label{action}
S=\int \mathrm{d}^{4} x \sqrt{-g}\left[\frac{R}{2}+\lambda (X+c~Y-1)\right],
\ee
where $\la$ enforces the mimetic constraint  \eqref{mc2}.
We will use the Lagrange  multiplier formulation in the left part of this paper. 

The equation of motion  for the metric $g_{\mu\nu}$ is 
\be
G_{\mu\nu}=-T_{\mu\nu},
\ee
 where the effective energy-momentum tensor for the double mimetic fields is 
\be\label{Tuv}
T_{\mu\nu}=2\lambda (\phi_{,\mu}\phi_{,\nu}+c~ \psi_{,\mu}\psi_{,\nu}).
\ee
In addition, we have the equations of motion for the two scalar fields
\be
\left(\sqrt{-g} \lambda \phi^{,\mu}\right)_{, \mu}=0, \quad\left(\sqrt{-g} \lambda  \psi^{, \mu}\right)_{, \mu}=0,
\ee
which are nothing but  the conservation equation of Noether currents associated with the shift symmetries. The Noether charges are correspondingly
\be
Q_{\phi}=\int\td^3x \sqrt{-g} \lambda \phi^{,0},\quad Q_{\psi}=\int\td^3x \sqrt{-g} \lambda \psi^{,0}.
\ee

\section{Cosmological applications}\label{sec:cosmology}
In this section, we study the implications of the two-field mimetic gravity model to cosmology. We first discuss the evolution of the Friedmann-Robertson-Walker (FRW) background, and then the linear perturbations.
\subsection{background}
 In the Arnowitt-Deser-Misner (ADM) formalism,  the metric can be decomposed as
\be
d s^{2}=N^{2} d t^{2}-h_{i j}\left(d x^{i}+N^{i} d t\right)\left(d x^{j}+N^{j} d t\right)
\ee
where $N$ is the lapse function, $N^i$ is the shift vector and $h_{ij}$ is the spatial metric. For the spatially flat FRW background, one have 
\be
N=1, ~N^i=0,~h_{ij}=a^2\delta_{ij},
\ee
where $a$ is the scale factor and we have used the cosmic time. Besides, we assume $\phi$,$\psi$, and $\lambda$ are functions only of time. The background equations are then given by
\bea
3H^2&=2\lambda,\\
3H^2+2\dot{H}&=0,\\
\dot{\phi}^2+c~\dot{\psi}^2&=1,
\eea
and
\be\label{c12}
\lambda a^{3} \dot{\phi}=c_{1}, \quad \lambda a^{3} \dot{\psi}=c_{2},
\ee
where $c_1$ and $c_2$ are just integration constants.  By employing the mimetic constraint, $\dot{\phi}^2+c~\dot{\psi}^2=1$, Eq.~\eqref{c12} implies
\be\label{lambda}
\lambda \propto a^{-3}~,\quad\dot{\phi}=\rm{const}~,\quad \dot{\psi}=\rm{const}.
\ee
Furthermore, one can read out the effective energy density and pressure of the mimetic fields: $\rho=2\la,~p=0$, from the energy-moment tensor \eqref{Tuv}. This yields $\rho \propto a^{-3}$ combined with Eq.~\eqref{lambda}. Therefore these two mimetic fields indeed behave like the dark mater in the universe, this is true at least at the background level. 

\subsection{adiabatic and entropy decomposition}
The projections of multiple canonical scalar fields into the adiabatic and entropy modes in the field space have been studied in Refs.~\cite{ Gordon:2000hv,Cai:2013kja,Gao:2008dt,Gao:2012uq}, and the generalizations to the non-standard scalar fields were  discussed in Ref.~\cite{Li:2014yla}.  In the two-field mimetic model considered here the  adiabatic mode $\sigma$, which represents the evolving path along the background trajectory, is given by 
\be
\dot{\sigma}=\sqrt{\dot{\phi}^2+c~\dot{\psi}^2}.
\ee
Using mimetic constraint, one immediately  has
\be
\dot{\sigma}=1.
\ee

When the perturbations are included, one can decompose the fluctuations of the two scalar fields into the adiabatic $\delta \sigma$  and entropy fluctuations $\delta s$ as
\bea\label{decomposition}
 \delta \sigma &=\frac{\dot{\phi}}{\sqrt{\dot{\phi}^{2}+c~ \dot{\psi}^{2}}} \delta \phi+\frac{c~\dot{\psi}}{\sqrt{\dot{\phi}^{2}+c~ \dot{\psi}^{2}}} \delta \psi, \\
 \delta s &=-~ \frac{\dot{\psi}}{\sqrt{\dot{\phi}^{2}+c~ \dot{\psi}^{2}}} \delta \phi+\frac{\dot{\phi}}{\sqrt{\dot{\phi}^{2}+c~ \dot{\psi}^{2}}} \delta \psi.
 \eea
One can see that $\delta \sigma$ represents the perturbation along the background trajectory  
and $\delta s$  is the perturbation orthogonal to the adiabatic direction. We should mention that the definition of ``orthogonal" here depends on the metric of the field space. If both mimetic fields have the same sign, i.e., $c=1$, the metric of field space is Euclidean; otherwise, $c=-1$, the metric is Minkowskian. 

Note that the definition of the rotation angle $\theta$ from the space of $(\delta\phi, ~\delta\psi)$ to the space of $(\delta\sigma,~\delta s) $\cite{Gordon:2000hv} does not work for the case $c=-1$.  For this case one should define something like rapidity $\eta$  as
\be
\cosh \eta = \frac{\dot{\phi}}{\sqrt{\dot{\phi}^{2}-~ \dot{\psi}^{2}}}, \quad \sinh \eta = \frac{\dot{\psi}}{\sqrt{\dot{\phi}^{2}-\dot{\psi}^{2}}},
\ee
and adiabatic $\delta \sigma$  and entropy fluctuations $\delta s$ are given by
\begin{align} 
\delta \sigma &=(\cosh \eta) \delta \phi-(\sinh \eta) \delta \psi, \\ 
\delta s &=-(\sinh \eta) \delta \phi+(\cosh \eta) \delta \psi.
\end{align}
Anyway, Eq.~\eqref{decomposition} is general and can apply to  both cases of $c=\pm1$. Hence, we will use the definition Eq.~\eqref{decomposition}  in the subsequent discussions on the perturbations.

\subsection{perturbation analysis}
In this subsection, we revisit on  the linear perturbation theory of two-field mimetic gravity model Eq.~\eqref{action}. The ADM decomposition of the action is
\begin{align}\label{ADM}
	S=\int\td^4x N\sqrt{h}\bigg\{&\frac{1}{2}(-^{(3)}R+K^{ij}K_{ij}-K^2)+\la\bigg[\frac{(\dot{\phi}-\phi_{,i}N^i)^2}{N^2}-h^{ij}\phi_{,i}\phi_{,j}\nn
	&+c\frac{(\dot{\psi}-\psi_{,i}N^i)^2}{N^2}-c~ h^{ij}\psi_{,i}\psi_{,j}-1\bigg]\bigg\}
\end{align}
where $c=\pm1$, $^{(3)}R$ is the spatial Ricci scalar associated to the metric $h_{ij}$, and $K_{i j}=(\dot{h}_{i j}-D_{i} N_{j}-D_{j} N_{i})/2N$ is the extrinsic curvature. 
As for the tensor perturbation it is trivial and just the same as Einstein's general relativity, we only consider the scalar perturbations in this paper. For convenience, we perform the perturbation analysis in comoving  gauge $\delta\sigma=0$. Expanding the ADM variables to linear order, one has
\be
N=1+A,\quad N^i=\pa_{i}B,\quad h_{ij}=a^2e^{2\mathcal{R}}\delta_{ij}~,
\ee
Where $\mathcal{R}$ is the comoving curvature perturbation.

The action  for the scalar perturbations  takes the form
\be
S^{(2)}_{com}=\int\td^4x \mathcal{L}^{(2)}_{com},
\ee
with
\begin{align}
 \frac{\mathcal{L}^{(2)}_{com}}{a^3}=&-3 \dot{\mathcal{R}}^{2}-18 H \mathcal{R} \dot{\mathcal{R}}-\frac{27}{2} H^{2} \mathcal{R}^{2}+6 H A \dot{\mathcal{R}}+9 H^{2} A \mathcal{R}-3 H^{2} A^{2} \nn
  &-\frac{1}{a^{2}}\left[(\partial \mathcal{R})^{2}+2\left(A+\mathcal{R}\right) \partial^{2} \mathcal{R}\right] +6 H \partial_{i} \mathcal{R} \partial_{i} B+2 \dot{\mathcal{R}} \partial^{2} B-2 H\left(A-3 \mathcal{R}\right) \partial^{2} B \nn
&+\lambda\left[c\l(\delta \dot{s}^{2}-\frac{1}{a^2}(\pa \delta s)^{~2}\r)+A^{2}-6 A \mathcal{R}\right]-2 \delta\lambda A~, \nonumber
 \end{align}
where the last line denotes the contribution from mimetic term and the others represent the contribution of the Einstein-Hilbert term.

After doing some integrations by parts and considering the background  EOMs,  we obtain the quadratic Lagrange density in the Fourier space
\be\label{L0}
 \mathcal{L}_{\mathrm{com}}^{(2)}=\LM_{\d s}^{(2)}+\LM_{\RM}^{(2)}
\ee
with 
\be\label{Ls}
 \LM_{\d s}^{(2)}= \frac{3}{2} c~a^{3} H^{2} (\delta \dot{s}^{2}-\frac{k^{2}}{a^2} \delta s^{2}), 
 \ee
 and 
 \begin{align}\label{LR}
\LM_{\RM}^{(2)} =&-3 a^{3} \dot{\mathcal{R}}^{2}+2 a^{3}\left(3 H A-k^{2} B\right) \dot{\mathcal{R}}+a k^{2} \mathcal{R}^{2}+2 a k^{2} A R \nn
 &-\frac{3}{2} a^{3} H^{2} A^{2}+2 a^{3} H k^{2} B A+2 a^{3} \delta\lambda A.
 \end{align}
We can see the  entropy perturbation is decoupled from the adiabatic curvature perturbation  at the linear perturbation level. 

From the Lagrange density \eqref{Ls}, it can be seen that the entropy mode propagates with the speed of unity, and whether the entropy perturbation is pathological exactly depends on $c$. In the case of $c=1$, where the kinetic terms of the original mimetic fields have the same sign in the constraint equation, the entropy perturbation is healthy, behaves like a massless canonical field coupled to the background.  However, in the case of $c=-1$ where the kinetic terms of $\phi$ and $\psi$ have opposite signs, the Lagrange density of the entropy perturbation $\delta s$ has a wrong sign. This means $\delta s$ is a ghost in this case and always suffers from the problem of quantum instability.

The momentum conjugated to $\d s$ is $\pi_{\d s}=3 c a^3H^2\dot{\d s}$ and  the Hamiltonian density for entropy perturbation is
\be \label{H1}
	\HM_{\d s}^{(2)}=c\l(\frac{\pi_{\d s}^2}{6a^3H^2}+\frac{3}{2}aH^2k^2{\d s}^2\r).
\ee

The variation of the perturbation action with respect to $\delta\la$  yields a constraint equation
\be\label{Acom} 
	A=0.
\ee
Then substituting  the constraint into the quadratic Lagrangian \eqref{LR}, we obtain the reduced Lagrangian
\be \label{LcomBR}
 \LM_{R}^{(2)}=-3a^3\dot{\RM}^2-2a^3k^2B\dot{\RM}+ak^2\RM^2
\ee
Variations with respect to $B$ and $\RM$ respectively yield the following two equations of motion
\be\label{BandR} 
	\dot{\RM}=0,~~\frac{d}{dt}(a^3B)+a\RM=0.
\ee
Here we lay out our difference from Ref.~\cite{Firouzjahi:2018xob}. The authors of \cite{Firouzjahi:2018xob} first got the equation $\dot{\RM}=0$ and then substituted it back into the Lagrangian (\ref{LcomBR}), finally through continuous variation with respect to $ \RM$ they obtained a vanishing curvature perturbation $\RM=0$.  So it was concluded in Ref.~\cite{Firouzjahi:2018xob} that there is only one scalar DOF (the entropy mode) in this two-field mimetic gravity model. In our opinion, we should not substitute $\dot{\RM}=0$ 
 into the Lagrangian \eqref{LcomBR} before the variation to $\RM$,  because it involves time derivative of $\RM$ and  thus can not be treated as an algebraic constraint equation of $\RM$. We think we should do the variations to $B$ and $\RM$ simultaneously as we have done above to get the EOMs (\ref{BandR}), which imply that we need two initial conditions to work out time evolutions of variables $B$ and $\RM$. Thus our viewpoint is that generally the adiabatic perturbation $\RM$ does not vanish in the two-field mimetic gravity model, so totally there are two scalar DOFs instead of one in this model, even though the extra DOF (the adiabatic mode) does not propagate.  

To confirm our result got above, we should perform the Hamiltonian analysis of  the quadratic Lagrangian density \eqref{LcomBR}.  Because it does not contain time derivative of $B$, one soon has the primary constraint $\pi_B=0$. The momentum associated with $\RM$ is $\pi_{\RM}=-6a^3\dot{\RM}-2a^3k^2B$ and the Hamiltonian for adiabatic perturbation is
\be 
\HM_{R}^{(2)}=-\frac{(\pi_\RM+2a^3k^2B)^2}{12a^3}-ak^2\RM^2+v\pi_B.
\ee
In terms of the primary constraint $\pi_B\approx0$ and the consistency condition $\dot{\pi}_B\approx 0$ we obtain a secondary constraint 
$$\pi_{\RM}+2a^3k^2B\approx0.$$ 
After imposing these two constraints, we have the reduced canonical phase space and  reduced Hamiltonian for the adiabatic perturbation,  
\be\label{H2} 
	\HM^{(2)}_{\RM}=-ak^2\RM^2.
\ee
One can check  that the canonical equations 
\be\label{RandPi} 
	\dot{\RM}=\{\RM,\HM^{(2)}_{\RM}\}=0,~~\dot{\pi}_\RM=\{\pi_\RM,\HM^{(2)}_{\RM}\}=2ak^2\RM
\ee
is equivelent to the EOM \eqref{BandR}. It is clear now that the adiabatic mode contributes one physical DOF, different from Ref.~\cite{Firouzjahi:2018xob}. 

Therefore, the total reduced Hamiltonian for scalar perturbations is
\be 
  \HM^{(2)}=c\l(\frac{\pi_{\d s}^2}{6a^3H^2}+\frac{3}{2}aH^2k^2{\d s}^2\r)-ak^2\RM^2.
\ee
Finally we have demonstrated at the linear perturbation level that there are two scalar DOFs in the two-field mimetic gravity \eqref{action}. Together with two tensor modes, the total DOFs of this model is four. In the next section, we will check the number of DOFs of this model at the full non-linear level in terms of the Hamiltonian method. 

\section{Hamiltonian analysis}\label{sec:Hamiltonian}
To identify the degrees of freedom  of the non-linear system \eqref{action}, we shall perform the  Hamiltonian analysis without perturbative  expansion. The action we start with is
\be 
	S=\int\td^4x\sqrt{g}\left[\frac{R}{2}+\la(g^{\mu\nu}\phi_{,\mu}\phi_{,\nu}+c ~g^{\mu\nu}\psi_\mu\psi_{,\nu}-1)\right],\nonumber
\ee
where $c=\pm1 $. In the ADM formalism, the action takes the form \eqref{ADM}. 

As the time derivative of $N$, $N^i$ and $\la$ are not involved in \eqref{ADM}, we have five primary constraints
$$\pi_N\approx0,~~~ \pi_i\approx0,~~~\pi_\la\approx0.$$ Other conjugate momentum are
\begin{align} 
\pi^{ij}&=\frac{\partial \LM}{\partial \dot{h_{ij}}}=\frac{\sqrt{h}}{2}(K^{ij}-h^{ij}K),\\
\pi_\phi&=2\la\frac{\sqrt{h}}{N}(\dot{\phi}-\phi_{,i}N^i),~\pi_\psi=2c\la\frac{\sqrt{h}}{N}(\dot{\psi}-\psi_{,i}N^i).\nonumber
\end{align}
Constructing the total Hamiltonian from the standard definition \cite{Bojowald}, we have
\be 
	H_T=\int\td^3x[N\HM+N^i\HM_i+v^N\pi_N+v^i\pi_i+v^\la\pi_\la],
\ee
where
\begin{align} 
 	\HM&=\HM_g+\HM_m\nn
	&=\sqrt{h}\left[\frac{^{(3)}R}{2}+\frac{2\pi^{ij}\pi_{ij}-\pi^2}{h}\right]+\sqrt{h}\left[\frac{\pi_\phi^2+\pi_\psi^2/c}{4\la h}+\la(h^{ij}\phi_{,i}\phi_{,j}+ c~h^{ij}\psi_{,i}\psi_{,j}+1)\right],\\
 	\HM_i&=\HM_{gi}+\HM_{mi}\nn
	&=-2\sqrt{h}\left(\frac{\pi^j_i}{\sqrt{h}}\right)_{|j}+\pi_\phi\phi_{,i}+\pi_\psi\psi_{,i}~.
\end{align}

The time evolution of $\Phi_1=\pi_\la\approx0$ leads to the mimetic constraint (secondary constraint)
\be 
  \Phi_2=-\frac{\pi_\phi^2+\pi_\psi^2/c}{4\la^2 h}+h^{ij}\phi_{,i}\phi_{,j}+ c~h^{ij}\psi_{,i}\psi_{,j}+1\approx0.
\ee
With the constraints $\Phi_1\approx0$ and $\Phi_2\approx0$, one can eliminate $\la$ and $\pi_\la$ to reduce the dimension of phase space and obtain the reduced Hamiltonian 
\be 
	H_{R}=\int\td^3x[N\HM_R+N^i\HM_i+v^N\pi_N+v^i\pi_i],
\ee
where
\begin{align}
 \HM_R=\sqrt{h}\bigg[\frac{^{(3)}R}{2}+\frac{2\pi^{ij}\pi_{ij}-\pi^2}{h}+\sqrt{\frac{(\pi_\phi^2+\pi_\psi^2/c)(h^{ij}\phi_{,i}\phi_{,j}+ c h^{ij}\psi_{,i}\psi_{,j}+1)}{h}}~\bigg],
\end{align}
which has the form similar to that of the original one-field mimetic gravity model \cite{Malaeb:2014vua}. Now we have 12 canonical conjugate  pairs for the reduced phase space, ten pairs from metric as in general relativity and two pairs from the mimetic scalar fields.  

The rest four primary constraints $\pi_N\approx0,~\pi_i\approx0 $ yield four secondary constraints 
\be 
  \HM_R\approx0,~~\HM_i\approx0~.
\ee
One can write the constraints in smeared form \cite{Bojowald}
\be 
	H[N]=\int\td^3x N\HM_R,\quad D[\vec{N}]=\int\td^3x N^i\HM_i~,
\ee
which is the generator of the spacetime diffeomorphism transformations. In order to test whether $\HM_R, \HM_i$ are first-class, one should work out the commutators among them.
The detailed calculations of the key Poisson brackets related to the Hamiltonian constraint $\HM_R$ and momentum constraint $\HM_i$ are straightforward and similar to what have done in the single field mimetic gravity model, one can refer to Ref.~\cite{Malaeb:2014vua} for more details, here we just give some critical steps. First we compute the commutator of momentum constraint with itself 
\begin{align}
\{D[\vec{M}],D[\vec{N}]\}=&\{D_g[\vec{M}],D_g[\vec{N}]\}+\{D_m[\vec{M}],D_m[\vec{N}]\}\nn
=&D_g[\LM_{\vec{M}}\vec{N}]+D_m[\LM_{\vec{M}}\vec{N}]=D[\LM_{\vec{M}}\vec{N}],
\end{align}
where $\LM_{\vec{M}}$ is the Lie derivative along $\vec{M}$. Then we compute the commutator of momentum constraint with the Hamiltonian constraint
\begin{align}
\{D[\vec{M}],H[N]\}=&\{D_g[\vec{M}],H_g[N]\}+\{D_g[\vec{M}]+D_m[\vec{M}],H_m[N]\}\nn
=&H_g[\LM_{\vec{M}}N]+H_m[\LM_{\vec{M}}N]=H[\LM_{\vec{M}}N],
\end{align}
and   the commutator of Hamiltonian constraint with itself 
\begin{align}
\{H[M],H[N]\}=&\{H_g[M],H_g[N]\}+\{H_m[M],H_m[N]\}\nn
=&D_g[h^{ij}(M\partial_jN-N\partial_jM)]+D_m[h^{ij}(M\partial_jN-N\partial_jM)]\nn
=&D[h^{ij}(M\partial_jN-N\partial_jM)].
\end{align}

The above Poisson algebra shows  all the commutators vanish on the constraint surface.\footnote{Because all the $H$ and $D$ are integrates of $\HM_R$ or $\HM_i$, which vanish on the constraint surface.} In addition, $\HM_R$ and $\HM_i$ do not rely on $N$ and $N^i$, so $\pi_N$ and $\pi_i$ commutate with all the constraints. In all,  the $8$  constraints $\{\pi_N, \pi_i, \HM_R, \HM_i\}$ are all first-class in the reduced Hamiltonian system. According to definition of DOFs by Dirac \cite{Dirac}, our system have
$(2\times12-2\times8)/2=4$ DOFs. This is consistent with our analysis of the linear perturbation in the previous section.  In the cosmological perturbation theory, the 4 DOFs correspond to one adiabatic mode, one entropy mode and two tensor modes respectively.

\section{conclusion} \label{sec:discussion}

In summary, we revisited in this paper the problem of DOFs in the two-field mimetic gravity model with shift symmetries, which was recently studied in Ref.~\cite{Firouzjahi:2018xob}. With quadratic Lagrangian and Hamiltonian formulations, our analyses on the linear perturbations around the homogeneous cosmological background showed that there are two scalar DOFs in this model, different from the result got in  Ref.~\cite{Firouzjahi:2018xob}. These two scalar DOFs correspond to the adiabatic and entropy perturbations respectively. The former does not propagate, same as in the model of  original single-field mimetic model. The entropy perturbation decouples from the adiabatic one and behaves like a massless scalar field, propagating in the homogeneous background. We further confirmed our result by analyzing the full non-linear theory in terms of the Hamiltonian formulation. In addition, we also considered the case where the kinetic terms of the two mimetic fields have opposite signs in the mimetic constraint equation. The quadratic Lagrangian show that, in this case, the entropy mode is a ghost. 

We point out here that the behavior of two-field mimetic gravity model \eqref{action} has many similarities to the original single-field mimetic theory. First,  the setup still mimics the roles of dark matter at the background level. Second, the comoving curvature is frozen $\dot{\RM}=0$. Therefore, if one focus only on the adiabatic curvature perturbation or the entropy perturbation is not excited at the beginning, two-field mimetic cosmology will be exactly the same as the single-field mimetic scenario at both the background and perturbation level. This implies that two-field mimetic gravity also suffer the same instability  issues which exist in single-field mimetic scenario, e.g. the Ostrogradski instability \cite{Takahashi:2017pje} at the presence of extra matter. 
To avoid these pathologies, some modifications and extensions should be made. 

If naively increasing the number of  mimetic scalar fields, one could expect that more entropy modes  will appear correspondingly while the adiabatic curvature perturbation is left unchanged and thus the Ostragradski  instability in the presence of a matter field remains. Besides, it is known that an inclusion of higher derivatives of the mimetic field in the case of single-field mimetic gravity can make the curvature perturbation propagating and resolve the  Ostragradski  instability in the presence of  matter field,  but unfortunately resulting in ghost or gradient instabilities.  For the two-field case, the situation may be similar. Moreover,  as there is more than one scalar field, the construction of higher derivative interaction terms  can be more general and complex, such as $f(X,Z,\Box\phi,\Box\varphi)$ which  respect shift symmetry and can't be simplified by employing the two-field mimetic constraint. 
Furthermore, one can consider  two-field mimetic models with  other  solutions for $A$ different from the form \eqref{cd2}. We hope to come back to study these topics soon.

\acknowledgments
This work is funded in part by  NSFC under Grant No. 11847239, 11961131007, 11421303, 11653002, 11422543 and the CAS Key Laboratory for Research in Galaxies and Cosmology, Chinese Academy of Science (No. 18010203).



\begin{thebibliography}{99}

\bibitem{Chamseddine:2013kea}
  A.~H.~Chamseddine and V.~Mukhanov,
  JHEP {\bf 1311} (2013) 135
  [arXiv:1308.5410 [astro-ph.CO]].

\bibitem{Golovnev:2013jxa}
  A.~Golovnev,
  Phys.\ Lett.\ B {\bf 728} (2014) 39
  [arXiv:1310.2790 [gr-qc]].

\bibitem{Chamseddine:2014vna}
  A.~H.~Chamseddine, V.~Mukhanov and A.~Vikman,
  JCAP {\bf 1406} (2014) 017
  [arXiv:1403.3961 [astro-ph.CO]].

\bibitem{Lim:2010yk}
E.~A.~Lim, I.~Sawicki and A.~Vikman,
  JCAP {\bf 1005} (2010) 012
  [arXiv:1003.5751 [astro-ph.CO]].
\bibitem{Saadi:2014jfa}
  H.~Saadi,
  Eur.\ Phys.\ J.\ C {\bf 76} (2016) no.1,  14
  [arXiv:1411.4531 [gr-qc]].


\bibitem{Mirzagholi:2014ifa}
  L.~Mirzagholi and A.~Vikman,
  JCAP {\bf 1506} (2015) 028
  [arXiv:1412.7136 [gr-qc]].

 \bibitem{Matsumoto:2015wja}
  J.~Matsumoto, S.~D.~Odintsov and S.~V.~Sushkov,
  Phys.\ Rev.\ D {\bf 91} (2015) no.6,  064062
  [arXiv:1501.02149 [gr-qc]].

\bibitem{Ramazanov:2015pha}
  S.~Ramazanov,
  JCAP {\bf 1512} (2015) 007
  [arXiv:1507.00291 [gr-qc]].

\bibitem{Myrzakulov:2015kda}
  R.~Myrzakulov, L.~Sebastiani, S.~Vagnozzi and S.~Zerbini,
  Class.\ Quant.\ Grav.\  {\bf 33} (2016) no.12,  125005
  [arXiv:1510.02284 [gr-qc]].

\bibitem{Astashenok:2015qzw}
  A.~V.~Astashenok and S.~D.~Odintsov,
  Phys.\ Rev.\ D {\bf 94} (2016) no.6,  063008
  [arXiv:1512.07279 [gr-qc]].

\bibitem{Chamseddine:2016uyr}
  A.~H.~Chamseddine and V.~Mukhanov,
  JCAP {\bf 1602} (2016) no.02,  040
  [arXiv:1601.04941 [astro-ph.CO]].

\bibitem{Nojiri:2016vhu}
  S.~Nojiri, S.~D.~Odintsov and V.~K.~Oikonomou,
  Phys.\ Rev.\ D {\bf 94} (2016) no.10,  104050
  [arXiv:1608.07806 [gr-qc]].

\bibitem{Babichev:2016jzg}
  E.~Babichev and S.~Ramazanov,
  Phys.\ Rev.\ D {\bf 95} (2017) no.2,  024025
  [arXiv:1609.08580 [gr-qc]].

\bibitem{Chamseddine:2016ktu}
  A.~H.~Chamseddine and V.~Mukhanov,
  Eur.\ Phys.\ J.\ C {\bf 77} (2017) no.3,  183
  [arXiv:1612.05861 [gr-qc]].

\bibitem{Shen:2017rya}
  L.~Shen, Y.~Mou, Y.~Zheng and M.~Li,
  Chin.\ Phys.\ C {\bf 42} (2018) no.1,  015101
  [arXiv:1710.03945 [gr-qc]].

\bibitem{Abbassi:2018ywq}
  M.~H.~Abbassi, A.~Jozani and H.~R.~Sepangi,
  Phys.\ Rev.\ D {\bf 97} (2018) no.12,  123510
  [arXiv:1803.00209 [gr-qc]].

\bibitem{Casalino:2018tcd}
  A.~Casalino, M.~Rinaldi, L.~Sebastiani and S.~Vagnozzi,
  Phys.\ Dark Univ.\  {\bf 22} (2018) 108
  [arXiv:1803.02620 [gr-qc]].

\bibitem{Vagnozzi:2017ilo}
  S.~Vagnozzi,
  Class.\ Quant.\ Grav.\  {\bf 34} (2017) no.18,  185006
  [arXiv:1708.00603 [gr-qc]].

\bibitem{Dutta:2017fjw}
  J.~Dutta, W.~Khyllep, E.~N.~Saridakis, N.~Tamanini and S.~Vagnozzi,
  JCAP {\bf 1802} (2018) 041
  [arXiv:1711.07290 [gr-qc]].

\bibitem{Chamseddine:2019pux}
	A.~H.~Chamseddine, V.~Mukhanov and T.~B.~Russ,
	arXiv:1908.03498 [hep-th].

\bibitem{Nojiri:2014zqa}
  S.~Nojiri and S.~D.~Odintsov,
  Mod.\ Phys.\ Lett.\ A {\bf 29} (2014) no.40,  1450211
  [arXiv:1408.3561 [hep-th]].

\bibitem{Leon:2014yua}
  G.~Leon and E.~N.~Saridakis,
  JCAP {\bf 1504} (2015) no.04,  031
  [arXiv:1501.00488 [gr-qc]].

\bibitem{Astashenok:2015haa}
  A.~V.~Astashenok, S.~D.~Odintsov and V.~K.~Oikonomou,
  Class.\ Quant.\ Grav.\  {\bf 32} (2015) no.18,  185007
  [arXiv:1504.04861 [gr-qc]].

\bibitem{Myrzakulov:2015qaa}
  R.~Myrzakulov, L.~Sebastiani and S.~Vagnozzi,
  Eur.\ Phys.\ J.\ C {\bf 75} (2015) 444
  [arXiv:1504.07984 [gr-qc]].


\bibitem{Rabochaya:2015haa}
  Y.~Rabochaya and S.~Zerbini,
  Eur.\ Phys.\ J.\ C {\bf 76} (2016) no.2,  85
  [arXiv:1509.03720 [gr-qc]].


\bibitem{Arroja:2015yvd}
  F.~Arroja, N.~Bartolo, P.~Karmakar and S.~Matarrese,
  JCAP {\bf 1604} (2016) no.04,  042
  [arXiv:1512.09374 [gr-qc]].

\bibitem{Cognola:2016gjy}
  G.~Cognola, R.~Myrzakulov, L.~Sebastiani, S.~Vagnozzi and S.~Zerbini,
  Class.\ Quant.\ Grav.\  {\bf 33} (2016) no.22,  225014
  [arXiv:1601.00102 [gr-qc]].

 \bibitem{Nojiri:2016ppu}
  S.~Nojiri, S.~D.~Odintsov and V.~K.~Oikonomou,
  Class.\ Quant.\ Grav.\  {\bf 33} (2016) no.12,  125017
  [arXiv:1601.07057 [gr-qc]].


 \bibitem{Momeni:2014qta}
  D.~Momeni, A.~Altaibayeva and R.~Myrzakulov,
  Int.\ J.\ Geom.\ Meth.\ Mod.\ Phys.\  {\bf 11} (2014) 1450091
  [arXiv:1407.5662 [gr-qc]].

\bibitem{Chamseddine:2018qym}
  A.~H.~Chamseddine and V.~Mukhanov,
  JHEP {\bf 1806} (2018) 060
  [arXiv:1805.06283 [hep-th]].

\bibitem{Chamseddine:2018gqh}
  A.~H.~Chamseddine and V.~Mukhanov,
  JHEP {\bf 1806} (2018) 062
  [arXiv:1805.06598 [hep-th]].

\bibitem{Zhong:2017uhn}
  Y.~Zhong, Y.~Zhong, Y.~P.~Zhang and Y.~X.~Liu,
  Eur.\ Phys.\ J.\ C {\bf 78} (2018) no.1,  45
  [arXiv:1711.09413 [hep-th]].
  
  \bibitem{Zhong:2018fdq}
  Y.~Zhong, Y.~P.~Zhang, W.~D.~Guo and Y.~X.~Liu,
  JHEP {\bf 1904} (2019) 154
  [arXiv:1812.06453 [gr-qc]].
  
\bibitem{Chamseddine:2019gjh}
	A.~H.~Chamseddine, V.~Mukhanov and T.~B.~Russ,
	arXiv:1908.01717 [hep-th].
	
\bibitem{Liu:2017puc}
  D.~Langlois, H.~Liu, K.~Noui and E.~Wilson-Ewing,
  Class.\ Quant.\ Grav.\  {\bf 34} (2017) no.22,  225004
  [arXiv:1703.10812 [gr-qc]].
   
\bibitem{Malaeb:2014vua}
  O.~Malaeb,
  Phys.\ Rev.\ D {\bf 91} (2015) no.10,  103526
  [arXiv:1404.4195 [gr-qc]].

 \bibitem{Chaichian:2014qba}
  M.~Chaichian, J.~Kluson, M.~Oksanen and A.~Tureanu,
  JHEP {\bf 1412} (2014) 102
  [arXiv:1404.4008 [hep-th]].

\bibitem{Takahashi:2017pje}
  K.~Takahashi and T.~Kobayashi,
  JCAP {\bf 1711} (2017) no.11,  038
  [arXiv:1708.02951 [gr-qc]].

\bibitem{Zheng:2018cuc}
  Y.~Zheng and T.~Liu,
  arXiv:1810.03826 [gr-qc].

\bibitem{Ganz:2018mqi}
  A.~Ganz, P.~Karmakar, S.~Matarrese and D.~Sorokin,
  Phys.\ Rev.\ D {\bf 99} (2019) no.6,  064009
  [arXiv:1812.02667 [gr-qc]].  
  
\bibitem{Malaeb:2019rdl}
  O.~Malaeb and C.~Saghir,
  Eur.\ Phys.\ J.\ C {\bf 79} (2019) no.7,  584
  [arXiv:1901.06727 [gr-qc]].
  
  \bibitem{Ganz:2019vre}
  A.~Ganz, N.~Bartolo and S.~Matarrese,
  arXiv:1907.10301 [gr-qc].

\bibitem{Hammer:2015pcx}
  K.~Hammer and A.~Vikman,
  arXiv:1512.09118 [gr-qc].

\bibitem{Golovnev:2018icm}
  A.~Golovnev,
  Phys.\ Lett.\ B {\bf 779} (2018) 441
  [arXiv:1801.07958 [gr-qc]].

\bibitem{Langlois:2018jdg}
  D.~Langlois, M.~Mancarella, K.~Noui and F.~Vernizzi,
  JCAP {\bf 1902} (2019) 036
  [arXiv:1802.03394 [gr-qc]].

\bibitem{Firouzjahi:2018xob}
  H.~Firouzjahi, M.~A.~Gorji, S.~A.~Hosseini Mansoori, A.~Karami and T.~Rostami,
  JCAP {\bf 1811} (2018) no.11,  046
  [arXiv:1806.11472 [gr-qc]].
  
\bibitem{Gorji:2018okn}
  M.~A.~Gorji, S.~Mukohyama, H.~Firouzjahi and S.~A.~Hosseini Mansoori,
  JCAP {\bf 1808} (2018) no.08,  047
  [arXiv:1807.06335 [hep-th]].

\bibitem{Sebastiani:2016ras}
  L.~Sebastiani, S.~Vagnozzi and R.~Myrzakulov,
  Adv.\ High Energy Phys.\  {\bf 2017} (2017) 3156915
  [arXiv:1612.08661 [gr-qc]].
 \bibitem{Ijjas:2016pad}
  A.~Ijjas, J.~Ripley and P.~J.~Steinhardt,
  Phys.\ Lett.\ B {\bf 760} (2016) 132
  [arXiv:1604.08586 [gr-qc]].

\bibitem{Firouzjahi:2017txv}
  H.~Firouzjahi, M.~A.~Gorji and S.~A.~Hosseini Mansoori,
  JCAP {\bf 1707} (2017) 031
  [arXiv:1703.02923 [hep-th]].

\bibitem{Zheng:2017qfs}
  Y.~Zheng, L.~Shen, Y.~Mou and M.~Li,
  JCAP {\bf 1708} (2017) no.08,  040
  [arXiv:1704.06834 [gr-qc]].
  
\bibitem{Hirano:2017zox}
  S.~Hirano, S.~Nishi and T.~Kobayashi,
  JCAP {\bf 1707} (2017) no.07,  009
  [arXiv:1704.06031 [gr-qc]].
  

\bibitem{Gorji:2017cai}
  M.~A.~Gorji, S.~A.~Hosseini Mansoori and H.~Firouzjahi,
  JCAP {\bf 1801} (2018) no.01,  020
  [arXiv:1709.09988 [astro-ph.CO]].


\bibitem{Deruelle:2014zza}
  N.~Deruelle and J.~Rua,
  JCAP {\bf 1409} (2014) 002
  [arXiv:1407.0825 [gr-qc]].

\bibitem{Arroja:2015wpa}
  F.~Arroja, N.~Bartolo, P.~Karmakar and S.~Matarrese,
  JCAP {\bf 1509} (2015) 051
  [arXiv:1506.08575 [gr-qc]].

\bibitem{Domenech:2015tca}
  G.~Domenech, S.~Mukohyama, R.~Namba, A.~Naruko, R.~Saitou and Y.~Watanabe,
  Phys.\ Rev.\ D {\bf 92} (2015) no.8,  084027
  [arXiv:1507.05390 [hep-th]].

\bibitem{Bekenstein:1992pj}
  J.~D.~Bekenstein,
  Phys.\ Rev.\ D {\bf 48} (1993) 3641
  [gr-qc/9211017].

\bibitem{Takahashi:2017zgr}
  K.~Takahashi, H.~Motohashi, T.~Suyama and T.~Kobayashi,
  Phys.\ Rev.\ D {\bf 95} (2017) no.8,  084053
  doi:10.1103/PhysRevD.95.084053
  [arXiv:1702.01849 [gr-qc]].

\bibitem{Vikman:2017gxs}
  A.~Vikman,
  arXiv:1712.10311 [astro-ph.CO].

\bibitem{Gordon:2000hv}
  C.~Gordon, D.~Wands, B.~A.~Bassett and R.~Maartens,
  Phys.\ Rev.\ D {\bf 63} (2001) 023506
  [astro-ph/0009131].

\bibitem{Gao:2008dt}
  X.~Gao,
  JCAP {\bf 0806} (2008) 029
  [arXiv:0804.1055 [astro-ph]].

\bibitem{Gao:2012uq}
  X.~Gao, D.~Langlois and S.~Mizuno,
  JCAP {\bf 1210} (2012) 040
  [arXiv:1205.5275 [hep-th]].
      
\bibitem{Cai:2013kja}
  Y.~F.~Cai, E.~McDonough, F.~Duplessis and R.~H.~Brandenberger,
  JCAP {\bf 1310} (2013) 024
  [arXiv:1305.5259 [hep-th]].

\bibitem{Li:2014yla}
  M.~Li,
  Phys.\ Lett.\ B {\bf 741} (2015) 320
  [arXiv:1411.7626 [hep-th]].

\bibitem{Bojowald}
M.~Bojowald,  Canonical gravity and applications: cosmology, black holes, and quantum gravity. Cambridge University Press, 2010.

\bibitem{Dirac}
P. A. M. Dirac: Lectures on Quantum Mechanics, Yeshiva University Press, New York, (1964).

\end{thebibliography}
\end{document}